\documentclass[12pt]{iopart}

\expandafter\let\csname equation*\endcsname\relax

\expandafter\let\csname endequation*\endcsname\relax
\usepackage{amsmath}  
\usepackage{amsfonts} 
\usepackage{graphicx} 
\usepackage[normalem]{ulem}
\usepackage[table]{xcolor}

\begin{document}


\title{Matter waves and clocks do not observe uniform gravitational fields}
\author{Peter Asenbaum$^{1,2}$, Chris Overstreet$^{1,3}$, Mark A. Kasevich$^{1}$}
\address{$^1$ Department of Physics, Stanford University, Stanford, California 94305, USA}
\address{$^2$ Institute for Quantum Optics and Quantum Information (IQOQI) Vienna, Austrian Academy of Sciences, Boltzmanngasse 3, 1090 Vienna, Austria.}
\address{$^3$ Department of Physics \& Astronomy, Johns Hopkins University, Baltimore, Maryland 21218, USA}

\ead{peter.asenbaum@oeaw.ac.at}

\date{\today}

\begin{abstract}
In a uniform gravitational field, classical test objects fall universally. Any reference object or observer will fall in the same universal manner. Therefore, a uniform gravitational field cannot create dynamics between observers and classical test objects. The influence of a uniform gravitational field on matter waves and clocks, however, is described inconsistently throughout research and education. 
To illustrate, we discuss the behavior of a matter-wave interferometer and a clock redshift experiment in a uniform gravitational field.  As a consistent formulation of the equivalence principle implies, a uniform gravitational field has no observable influence on these systems and is physically equivalent to the absence of gravity.
\end{abstract}


\section{Introduction} 


Gravity is usually described in one of two ways: either as a Newtonian field defined on three-dimensional space (Newtonian gravity) or as the curvature of a four-dimensional spacetime (Newton-Cartan and, including relativistic effects, general relativity) \cite{Misner1973, Carroll2004a, Wald2010}. The Newtonian field has a magnitude at every point in space, whereas the curvature is a geometric feature of spacetime and is not observable at any one point. How can these two concepts be reconciled? 
As illustrated in this text, the answer is that the Newtonian field is not observable at any one point either, also when using matter waves or clocks. In other words, a uniform gravitational field (UGF)\footnote{For non-relativistic effects, we use the term ``UGF'' to refer to a Newtonian
  gravitational field that does not depend on the position. In section \ref{rel}, ``UGF'' refers to the position-dependent metric in Eq. \ref{metric}. Note that the corresponding curvature tensor vanishes for this
  metric.} is not observable.

Newtonian gravity, Newton-Cartan, and general relativity all satisfy the equivalence principle (EP). Modern formulations of the equivalence principle \cite{Pauli1921,Chryssomalakos2003,DiCasola2015} stress that gravity cannot influence measurement outcomes in any local 
experiment\footnote{A ``local'' experiment is an experiment in which all length scales and time scales are small enough that the effects of gravity gradients are below the measurement resolution.}, even in experiments involving quantum states \cite{Will2014}. 
However, this basic tenet seems to be violated in the discussion of gravitational phenomena in quantum systems, since matter wave interferometers and clock redshift measurements are commonly held as evidence for observable effects due to the interaction of a quantum system with a UGF \cite{Colella1975,Greenberger1983,Mannheim1998, Okon2011, Pikovski2015,  Nauenberg2016,Seveso2017,Emelyanov2022, Huggett2023,Pound1959,Chou2010}.
This tension is caused by the unfortunate use of observers or reference objects that are ``fixed'' in the UGF. 

In the Newtonian model of gravity \cite{Misner1973}, massive bodies create a gravitational field ${\bf G} ({\bf x})$ that induces the force ${\bf F}_G({\bf x})=m_G\,  {\bf G}({\bf x})$ on an object with gravitational mass $m_G$ and position ${\bf x}$. The inertial mass parameter $m_i$ connects a force acting on an object with its acceleration. We assume the inertial and the gravitational mass to be identical\footnote{The equality between inertial and gravitational mass is usually termed the {\it Galilean equivalence principle} \cite{Galileo1914}.} and set $m_G=m_i=m$ throughout the text. This equality has been tested experimentally to high precision \cite{Eotvos1890,Schlamminger2008,Asenbaum2020,Touboul2022}. With this equality, the gravitational field ${\bf G}({\bf x})$ acts universally on all massive objects.  In a small enough region around ${ x}_0$, the gravitational field is approximately uniform, ${\bf G}({\bf x}_0)={\bf a}_G$. In this UGF, every object and observer will fall with acceleration ${\bf a}_G$. Relative accelerations between objects vanish.
A UGF cannot create relative dynamics between observers and test objects because it acts in a universal manner. The emphasis here lies on the field being uniform between the observer and the test object. In contrast, gravitational field differences can cause dynamics. For instance, an apple falls toward the center of the Earth because of the gravitational field difference between the apple's position and the Earth's position. The field contribution from other gravitational sources, e.g. the sun, remains unobservable.

To be fixed in a UGF implies that non-gravitational forces ${\bf F}_{NG}=m\, {\bf a}_{NG}$ are applied to counteract the gravitational acceleration induced by a UGF. For a fixed object, it is then assumed that the applied force causes an exactly equal and opposite acceleration ${\bf a}_{NG} = -{\bf a}_G$ \footnote{This assumption is motivated by our everyday intuition: The gravitational acceleration between an apple falling from a tree and the Earth is well approximated by the non-gravitational acceleration of the surface of the earth.}. In this text, we lift this restriction to avoid confusion between non-gravitational forces and gravitational acceleration. The distinction we would like to stress here is that you can  {\it feel} non-gravitational interactions but not the UGF \cite{Landerer2020}.

The paper is organized as follows.  In Section II we analyze matter wave interferometers \cite{Colella1975,Kasevich1992, Peters1999, Karcher_2018}, and in Section III we analyze clock redshift experiments \cite{Pound1959,Chou2010,Mehlstaeubler2018,Bothwell2022}.
These experiments demonstrate that in standard physics, a uniform gravitational field is not observable.
As implied by the equivalence principle, all observable effects in a UGF are of non-gravitational origin and are independent of the magnitude and direction of the UGF.

\section{Matter-wave interferometer in UGF} 
\label{interferometer}
Matter-wave interferometers are composed of massive particles and some form of diffraction grating that puts each particle in a spatial quantum superposition. The spatial superposition is usually described by two well-separated wave packets that travel along classical trajectories known as the ``interferometer arms.'' To predict the interferometer phase, one has to keep track of the positions of the interferometer arms relative to the diffraction gratings and of the phase evolution of each wave packet as it propagates.
The two crucial points are: First, matter-wave interferometers can measure the relative acceleration between the gratings and the interferometer arms; a UGF does not cause relative accelerations. Second, a UGF does not create a propagation phase difference between the two wave packets \cite{Storey1994}.

Here we consider light-pulse atom interferometers \cite{Hogan2009}, but our analysis can be mapped directly to other matter-wave interferometers, e.g., by exchanging the wavelength of the light with the lattice constant of a crystal used for neutron diffraction \cite{Colella1975}. In a light-pulse atom interferometer, a cloud of cold atoms is diffracted by laser gratings made from counter-propagating laser beams \cite{Asenbaum2017}.

\begin{figure}[t]
	\begin{center}
		\includegraphics[width=0.9\columnwidth]{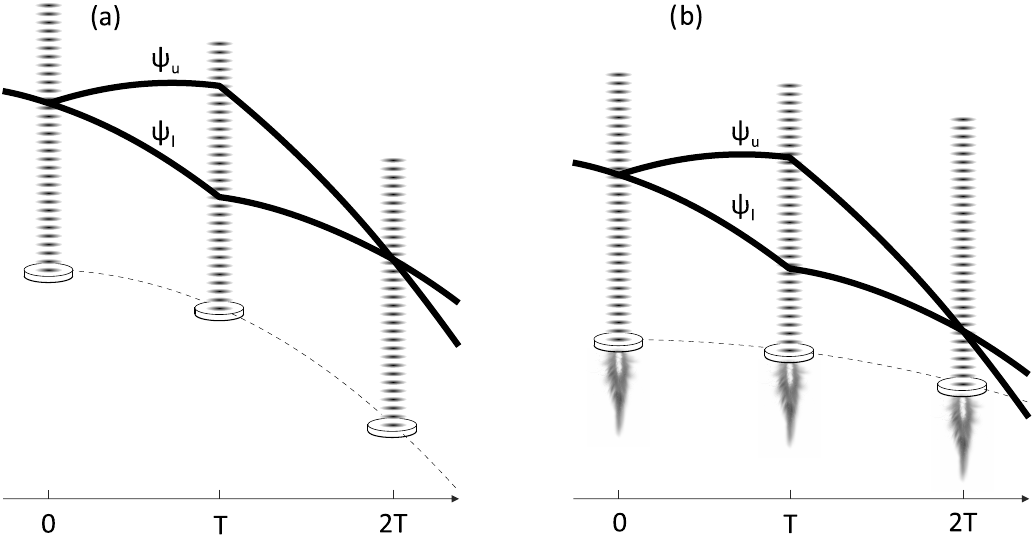}
        \caption{ Atom interferometer in a uniform gravitational field. a) The interferometer arms $\psi_u$ and $\psi_l$ fall universally between laser pulses, as does the mirror. The interferometer phase is zero. b) A gas propellant accelerates the retro-reflection mirror with acceleration $a_{NG}$, which influences the interferometer phase.}
		\label{fig:figure2}
	\end{center}
\end{figure}

The position of the laser grating is typically set by a retro-reflection mirror \footnote{The analysis in this section is restricted to one spatial dimension.}.  In a UGF, the calculation for predicting and interpreting the measured interferometer phase can be performed with the midpoint theorem \cite{Antoine2003}. Using this theorem allows for a straightforward interferometer phase calculation without losing accuracy for matter waves in a UGF. Later in this section, we will also discuss an approach using perturbation theory. The average position of the two interferometer arms is the midpoint trajectory. The distance between the midpoint and the mirror, which sets the position of the laser gratings, is denoted as $\bar{z}(t)$. The interferometer is composed of multiple laser gratings, which have the wave vectors $k_i$. These wave vectors encode the transferred momentum $\hbar k_i$ to the interferometer arms due to the diffraction process at the $i^\text{th}$ grating. In accordance with the midpoint theorem, the interferometer phase $\Phi$ is given by the sum 
\begin{equation}
\Phi=\Sigma_i \, k_i \, \bar{z}(t_i),
\end{equation}
where $t_i$ is the time of the diffraction by the $i^\text{th}$ laser grating. Without going into the specifics of the trajectories of the arms, this formula already highlights the fact that the phase depends only on the positions of the interferometer arms at the laser gratings. The momentum imparted by the grating interaction leads to a velocity change of $\hbar k/m$. In general, such a recoil velocity can cause an acceleration of the midpoint trajectory and therefore a measurable phase shift.

For inertial and gravitational sensing, it is desirable that the interferometer phase is insensitive to initial conditions and recoil effects from the laser interaction. Therefore, atom interferometers are operated in a closed and symmetric configuration, e.g. in a Mach-Zehnder interferometer. In such a configuration, the midpoint of the interferometer is not accelerated due to the laser recoil, and the interferometer arms overlap spatially at the end. For a Mach-Zehnder interferometer\cite{Kasevich1992}  with total duration $2T$, the phase $\Phi$ is given by
\begin{equation}
\Phi = k\, \tfrac{z_u (0) + z_l (0)}{2} - 2k\, \tfrac{z_u (T) + z_l (T)}{2}   + k\, \tfrac{z_u (2T) + z_l (2T)}{2} 
\end{equation}
where $z_u(t)$ and $z_l(t)$ are the positions of the upper and lower interferometer arms at time $t$, respectively. 

In Fig.~\ref{fig:figure2}(a), the mirror is freely falling, and the interaction with the laser light causes a negligible acceleration of the mirror. At the first interferometer pulse, the atoms are displaced by $z_0=z_u (0)=z_l (0)$ from the mirror and have no initial relative velocity. At the second pulse, the upper arm is displaced by $z_u(T)=z_0 +\hbar k T/(2m) $ and the lower arm by $z_l (T) = z_0-\hbar k T /(2m)$, where $m$ is the atomic mass.  At the third pulse, both paths overlap at a distance of $z_u (2T)=z_l (2T)=z_0$ from the mirror. Adding together the phase contributions, one obtains 
\begin{equation}
\Phi = 0 
\end{equation}
for the configuration of a freely falling mirror and freely falling atoms. A UGF has no observable effect on the atom interferometer.

If instead the mirror is non-gravitationally accelerated with acceleration $a_{NG}$ as shown in Fig.~\ref{fig:figure2}(b), the interferometer phase $\Phi$ reads

\begin{align}
\Phi = k z_0\, -\, & k\,( \overbrace{z_0 +\hbar k T/2m - a_{NG}T^2 /2}^{z_u(T)} +\overbrace{z_0 -\hbar k T/2m - a_{NG}T^2 /2}^{z_l(T)} ) \\ 
& + \frac{k}{2}\,( \overbrace{z_0 - 2a_{NG}T^2}^{z_u(2T)} +\overbrace{z_0 - 2a_{NG}T^2}^{z_l(2T)})  \nonumber
\end{align}
\begin{equation}
= - k a_{NG} T^2.
\label{result}
\end{equation}
In this case, the atom interferometer measures the phase shift $- k a_{NG}T^2$ induced by the non-gravitational acceleration of the mirror.  Once again, the UGF is not observed. 

The propagation phase of the wave packets along the interferometer arms is given by the classical action \cite{Storey1994}. The midpoint theorem takes advantage of the fact that the action difference between the arms is zero \cite{Roura2020}.  If instead one explicitly computes the action difference, the gravitational potential energy difference between arms gives rise to a phase term
\begin{equation}
\phi_{V_G}= -\frac{m}{\hbar}\int_0^{2T} \left[V_G(z_u)-V_G(z_l) \right] dt \label{potentialEqn}
\end{equation}
where $V_G(z)= -a_G (z-z_0)$ is the gravitational potential of the UGF.  On average, the upper arm is located higher by $\Delta z=\hbar k T /2 m$, so $\phi_{V_G}$ simplifies to 
\begin{equation}
\phi_{V_G} = \frac{m}{\hbar}\, a_G\, \Delta z\, 2 T = k a_G T^2. \label{wrongResult1}
\end{equation}
The phase term arising from the kinetic energy difference between the arms is given by 
\begin{equation}
\phi_{T} = \frac{m}{2\hbar} \int_0^{2T} \left[v_u(t)^2- v_l(t)^2\right] dt =-k a_G T^2,
\end{equation}
with $v_u (t)$ and $v_l(t)$ being the velocity along the upper and lower arm of the interferometer. The phase term $\phi_{V_G}$ from the potential energy difference is canceled exactly by the phase term $\phi_T$  from the kinetic energy difference \cite{Storey1994,Laemmerzahl1996, Antoine2003,Wolf2011, Overstreet2021}, so the UGF does not give rise to an observable phase shift.

In a different approach, the gravitational field is often treated as perturbing potential and the phase shift of the interferometer is calculated by using perturbation theory \cite{Storey1994,Ufrecht2020}. The appeal of this approach is that one can calculate the lowest order phase shift without the need to solve for the perturbed quantum state. In this calculation, the propagation phase shift is given just by the integral of the gravitational potential energy difference along the unperturbed interferometer paths, which is again equal to $\phi_{V_G}$. The phase shift term due to the laser interaction is given just by the phase shift of the lasers since the interferometer paths used for the calculation are undeflected \cite{Storey1994}. But as before, the phase of the lasers comes into the interferometer phase \footnote{See equation (91) in the tutorial by P. Storey and C. Cohen-Tannoudji \cite{Storey1994}.} and is determined by the position of the phase reference, the mirror. 
The mirror accelerates due to the UGF, and the laser phase shift is equal and opposite to the propagation phase shift. 
Without non-gravitational acceleration of the mirror, the interferometer phase shift due to a UGF is predicted to be zero by perturbation theory, in agreement with non-perturbative calculations.

Note that the interferometer phase in a UGF does not depend on the mass $m$ of the test particle. Empirically, how well do we know that the phase shifts of small quantum states do not depend on the mass? By sending two matter waves with different masses through the same interferometer, phase shifts due to the Earth's field that are proportional to the mass of the atom have been excluded by ten orders of magnitude \cite{Schlippert2014, Asenbaum2020}. Gravitational phase shifts proportional to the test particle mass only show up once the wavepacket separation becomes large in comparison to the distance to the gravitating source mass \cite{Overstreet2022}\footnote{Naturally, mass-dependent phase shifts can also arise in geometries where the
  non-gravitational momentum transfer or midpoint displacement is a function of
  the mass, e.g. in guided interferometers \cite{Xu2019} or recoil-sensitive
  interferometers. Such mass dependence does not indicate the presence of a
  gravitational effect.}.%

\section{Gravitational redshift in UGF}
\label{rel}

\begin{figure}
	\begin{center}
		\includegraphics[width= 0.9\columnwidth]{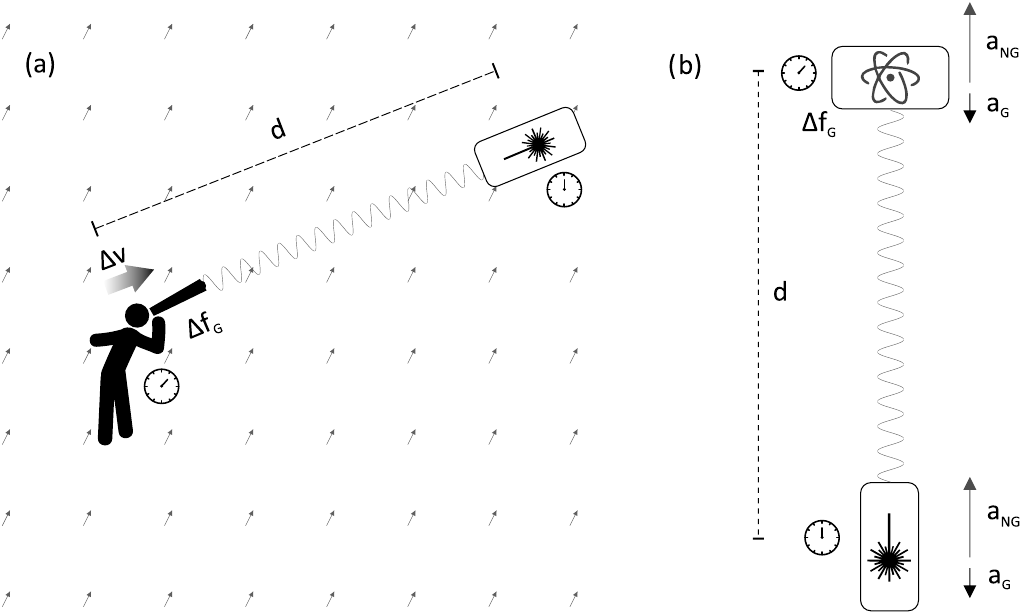}
        \caption{ Gravitational redshift in a uniform gravitational field (UGF). a) Detector and light source are freely falling in a UGF with distance $d$. While the light travels from the source to the detector, the detector continues to fall and gains additional velocity, which results in a Doppler shift $\Delta f_D$ of the observed frequency. In general relativity, a gravitational field causes clock rates to depend on position, leading to gravitational redshift $\Delta f_G$.  When the observer compares the Doppler-shifted light to his shifted clock, the Doppler shift cancels the gravitational redshift, and the UGF is not observed. b) The light source and the detector fall in the UGF with acceleration $a_G$ and are also accelerated non-gravitationally at a rate $a_{NG}$. The observed frequency shift is a function of $a_{NG}$ but not $a_{G}$.}
		\label{fig:figure3}
	\end{center}
\end{figure}

So far, we have not considered whether relativistic effects are predicted to be observable in a UGF. Let us assume that a light source in a UGF emits light with a certain frequency $f_0$. The light source is displaced by ${\bf d}$ from a detector with equal velocity. While the light travels toward the detector over a duration $|{\bf d}|/c$, the detector is falling with acceleration ${\bf a}_G$ and gains additional velocity $\Delta v ={\bf a}_G \cdot {\bf d}/c$ in the displacement direction, as shown in Fig.~\ref{fig:figure3}(a). Due to the Doppler effect, the received light is blue-shifted \cite{Radosz2009} from the emitted wavelength by the frequency $\Delta f_D$, where $\Delta f_D/f_0 ={\bf a}_G \cdot {\bf d}/c^2$ to lowest order in $\Delta v/c$. 

If it were possible to measure this blueshift of the light from the freely falling object, the EP formulation ``gravity is not locally observable" would be violated. 
For the UGF not to violate this EP formulation, there must be a compensating redshift $\Delta f_G$ induced by the UGF---the ``gravitational redshift.'' 

The gravitational redshift was originally presented as a consequence of energy conservation \cite{Einstein1911, Misner1973, Rindler2003},  as the photons travel along the Newtonian gravitational field.  In general relativity, a UGF can be simulated by the line element \cite{Rohrlich1963} 
\begin{equation}
     d \tau^2 = g_{\mu \nu}dx^\mu dx^\nu /c^2= (1- {\bf a}_G \cdot {\bf x}/c^2)^2 dt^2-d{\bf x}^2/c^2.
\label{metric}
\end{equation}
where $g_{\mu \nu}$ is the metric tensor. Usually, a coordinate transformation is performed to remove the dependence on ${\bf a}_G $ and to create a trivial situation without any gravitational dynamics. 
Here, we do not apply such a transformation to emphasize the difference between ${\bf a}_G $ and ${\bf a}_{NG} $.
The source clock and the detector clock are separated by {\bf d} and show time $\int d\tau_s$ and $\int d\tau_d$, respectively. For small velocities and small coordinate time interval $dt$, the clock time difference is $d\tau_d -d\tau_s={\bf a}_G \cdot {\bf d}/c^2 \, dt$. The clocks will measure frequencies differently by the gravitational redshift $\Delta f_G / f_0 = -{\bf a}_G \cdot {\bf d}/c^2$ to lowest order in $\Delta v/c$.

The Doppler shift from falling in the UGF and the gravitational redshift are equal and opposite, $\Delta f_G = -\Delta f_D$; they cancel out so that relativistic frequency shifts cannot be observed in freely falling local systems. Whereas  a gravitational redshift is not observable in a UGF, differences in the gravitational field can cause observable frequency shifts \cite{Delva2018,Herrmann2018,Do2019}.

Finally, we consider a configuration [Fig.~\ref{fig:figure3}(b)] in which a light source and a detector are non-gravitationally accelerated with ${\bf a}_{NG}$, 
as in the Pound-Rebka experiment \cite{Pound1959}. The acceleration 
of both the light source and the detector is given by ${\bf a}_{NG}+{\bf a}_{G}$. 
To lowest order in $\Delta v/c$, the total redshift $\Delta f$ is given by the sum of three terms:
\begin{equation}
    \Delta f = \Delta f_D^{NG} + \Delta f_D^{G} + \Delta f_G 
\end{equation}
where $\Delta f_D^{NG}/f_0 = {\bf a}_{NG}\cdot \mathbf{d}/c^2$ and $\Delta f_D^{G}/f_0 = {\bf a}_{G} \cdot \mathbf{d}/c^2$ are the Doppler shifts associated with the non-gravitational and gravitational acceleration, respectively, and $\Delta f_G$ is the gravitational redshift.  As before, $\Delta f_D^{G}$ and $\Delta f_G$ cancel out. 
The observable frequency shift is given solely by the Doppler shift associated with the non-gravitational acceleration.  
This is confirmed by experimental tests in which the measured redshift is consistent with the locally measured acceleration of the light source and the detector \cite{Pound1959} or a pair of precise optical clocks \cite{Takamoto2020} \footnote{Redshift is predicted to affect accelerated clock states in a spatial quantum superposition \cite{Pikovski2015,Roura2020}. The calculated dephasing arises from the non-gravitational acceleration that ``fixes'' the quantum states in a UGF. However, this effect is independent of the UGF, which does not cause an observable redshift \cite{Bonder2016}.}.

These results demonstrate the absence of an observable redshift from a UGF to their precisions. A local redshift test in free fall, where ${\bf a}_{NG} = 0$, should be possible in the near future.

\section{Conclusion}

We have considered the influence of a UGF on the physical observables associated with a quantum test mass and a pair of clocks.  In the absence of non-gravitational interactions, a UGF does not affect any observable quantities.  When non-gravitational interactions are introduced, the resulting changes in physical observables are caused by the non-gravitational interactions, not by the UGF. 

These conclusions are a direct confirmation of the equivalence principle, which states that gravity is not observable in local systems.  If there are no non-gravitational forces, any observer and any experiment in a UGF are in free fall, and all measurement results must be identical to what is obtained in the absence of gravitational sources.  The observation of nontrivial relative dynamics in a UGF indicates that non-gravitational forces are present.

The equivalence principle is often illustrated in a thought experiment comparing physical
effects in a UGF on Earth and in a rocket ship far from any gravitational source. Since a UGF is not observable, this thought experiment shows the equivalence of a UGF and empty space. Unfortunately, this equivalence is obscured by the use of non-gravitational forces to fix observers and to accelerate the rocket. While such a thought experiment can be useful for illustration, it should not be taken as the definition of the equivalence principle, as the thought experiment invokes a nonlocal comparison with a far-away experiment that is inaccessible in practice.  In contrast, local formulations of the equivalence principle \cite{DiCasola2015} are directly applicable to the prediction of experiments. \cite{Nauenberg2016}.

Gravity is a nonlocal phenomenon and is associated with a length scale. On Earth, the magnitude of the gravitational acceleration is typically $3\times 10^{-7}\,g$ for objects separated by 1 meter. The strength of Earth's gravity is not given by $g=9.8\, $m/s$^2$ in any local region.

\section*{Acknowledgements}
The authors thank Minjeong Kim and Joseph Curti for fruitful discussions.
This work was supported by the Vannevar Bush Faculty Fellowship Program. The work of C.O. was partially supported by the U.S. Department of Energy, Office of Science, National Quantum Information Science Research Centers, Superconducting Quantum Materials and Systems Center (SQMS) under contract number DE-AC02-07CH11359. The authors have no relevant financial or non-financial interests to disclose.
\\

\section*{Appendix 1:  Gravimetry}

Gravimetry experiments do not measure the local value of the gravitational field.  Instead, gravimeters use local measurements of non-gravitational (proper) acceleration, combined with nonlocal position information, to infer the gravitational field difference between two spatially separated points.

As an example, we consider a local gravimetric measurement at a distance $R$ from a gravitational source of mass $M$.  According to Newtonian mechanics, the gravitational field has magnitude $g = GM/R^2$ and points toward the source mass.  The measuring apparatus consists of a spring balance and a test mass.  We assume that the apparatus is small enough and its resolution low enough that it can be treated as a local system.  

If this measuring apparatus is placed at a distance $R$ from the source mass, then according to the equivalence principle, the spring balance and test mass will fall identically toward the source mass.  The measuring apparatus will read ``zero'' rather than $g$, and the local value of the gravitational field will not be observed.

If instead the spring balance is accelerated non-gravitationally at rate $a$ away from the source mass, then the measuring apparatus will read $a$.  In order to interpret this observation of non-gravitational acceleration as a gravimetric measurement, the experiment must be designed in such a way that the measuring apparatus remains at a fixed distance from the source mass.  For example, the proper acceleration of the spring balance can be actuated to keep the distance to the source mass constant, or the spring balance can be attached to a rigid body to constrain its position.  In any case, the experiment must incorporate nonlocal position information (namely, the relative position between the measuring apparatus and source mass), and position errors will induce uncertainty in the gravimetric interpretation of the measurement.

Once the position of the measuring apparatus is referenced to the source mass position, the observed quantity $a \approx g$ corresponds to the gravitational field {\it difference} between the positions of the measuring apparatus and source mass.  (In other words, if the measuring apparatus and source mass were falling in the approximately uniform gravitational field of another mass, the experiment would have no way to detect this.)  Even with the inclusion of nonlocal position information, gravimetry experiments do not measure the local gravitational field value but rather a gravitational tidal force.

\section*{Appendix 2:  Comparison with electromagnetism}

Unlike the gravitational field, the electromagnetic field can be measured locally.  To measure the electromagnetic field tensor at a point, it suffices to use two test particles, one of which is electrically charged and the other of which is neutral.  For example, the particles can be spatially overlapped, and the relative acceleration of the two particles can be observed.  By varying the initial velocity of the charged particle, all components of the electric and magnetic fields can be determined, provided the charge-to-mass ratio of the charged particle is known.

In contrast, an experiment like this one cannot be used to measure the local value of the gravitational field.  According to the equivalence principle, all particles have the same gravitational charge-to-mass ratio $m_G / m_i = 1$, so there is no ``gravitationally neutral'' particle that can serve as a reference.  Locally, a gravitational field does not induce relative acceleration between any two systems.

\section*{References}
\bibliographystyle{spiebib}
\bibliography{bib}

\end{document}